\documentclass[12pt,letterpaper]{article}
\usepackage[a4paper, total={7in, 10in}]{geometry}

\usepackage{graphicx}
\usepackage{helvet}
\usepackage{authblk}
\usepackage{hyperref}
\usepackage{amsmath} 
\usepackage{amssymb} 
\usepackage{orcidlink} 
\usepackage{booktabs}  
\usepackage{placeins}  
\usepackage[super,comma,sort&compress]  
   {natbib}
\usepackage[right]{lineno} \linenumbers

\makeatletter
\renewcommand{\maketitle}{\bgroup\setlength{\parindent}{0pt}
\begin{flushleft}
  \textbf{\@title}
  
  \@author
\end{flushleft}\egroup}
\makeatother

\title{Fine-tuning the ESM2 protein language model to understand the functional impact of missense variants}
\date{}

\author[1,2]{Ali Saadat}
\author[1,2,3,*]{Jacques Fellay}

\affil[1]{School of Life Sciences, Ecole Polytechnique Fédérale de Lausanne, Lausanne, Switzerland}
\affil[2]{Swiss Institute of Bioinformatics, Lausanne, Switzerland}
\affil[3]{Precision Medicine Unit, Biomedical Data Science Center, Lausanne University Hospital and University of Lausanne, Lausanne, Switzerland}

\affil[*]{Correspondence: jacques.fellay@epfl.ch}

\begin{document}

\maketitle

\section*{ABSTRACT}

Elucidating the functional effect of missense variants is of crucial importance, yet challenging. To understand the impact of such variants, we fine-tuned the ESM2 protein language model to classify 20 protein features at amino acid resolution. We used the resulting models to: 1) identify protein features that are enriched in either pathogenic or benign missense variants, 2) compare the characteristics of proteins with reference or alternate alleles to understand how missense variants affect protein functionality. We show that our model can be used to reclassify some variants of unknown significance. We also demonstrate the usage of our models for understanding the potential effect of variants on protein features.

\section*{KEYWORDS}


missense variant, mechanistic interpretation, protein language models, fine-tuning, token classification

\section{INTRODUCTION}

Recent advancements in sequencing technologies and bioinformatic analyses have significantly enhanced their utility in clinical settings, enabling more precise and comprehensive genetic diagnostics \cite{Bagger2024}. This progress has led to the generation of vast amounts of clinical-grade, personal genetic data, providing unprecedented opportunities to uncover the genetic basis of diseases. However, this surge in data also brings substantial challenges, particularly in the interpretation of variants of uncertain significance (VUS). Many putatively deleterious variants identified in the coding regions of the genome are missense variants, which can alter protein function by substituting one amino acid for another \cite{Chen2023}. Accurately determining the clinical significance and functional impact of these variants remains a formidable task \cite{Miosge2015}.

In the context of diagnostic genetic testing, the American College of Medical Genetics and Genomics (ACMG) and the Association for Molecular Pathology (AMP) guidelines are widely employed for sequence variant interpretation \cite{Richards2015}. These guidelines propose a standardized framework that integrates diverse types of evidence, including population data \cite{Chen2023,Auton2015}, computational predictions \cite{Cheng2023,Ioannidis2016}, and functional studies. One of the criteria in these guidelines, PM1 (moderate evidence of pathogenicity), considers whether a missense variant is located in a mutational hotspot or a critical, well-established functional domain (e.g., the active site of an enzyme) that lacks benign variation \cite{Richards2015}. While this criterion provides valuable guidance, identifying such regions poses significant challenges. It requires robust annotation of functional domains and systematic quantification of pathogenic and benign enrichment \cite{Harrison2019}.

Previous studies have attempted to address this challenge using gene-disease databases such as ClinVar \cite{Landrum2013} combined with population frequency data \cite{Chen2023,Auton2015}. These efforts have identified regions enriched with pathogenic variants and provided insights into specific proteins \cite{Quinodoz2022,Iqbal2020,PrezPalma2019}. However, these analyses have been limited by the incomplete annotation of the human proteome, leaving many proteins and functional regions unexplored. Furthermore, little attention has been paid to understanding the mechanistic impact of missense variants on specific protein features, which could provide deeper insights into their pathogenicity.

\begin{figure}[h]
    \centering
    \includegraphics[width=0.8\textwidth]{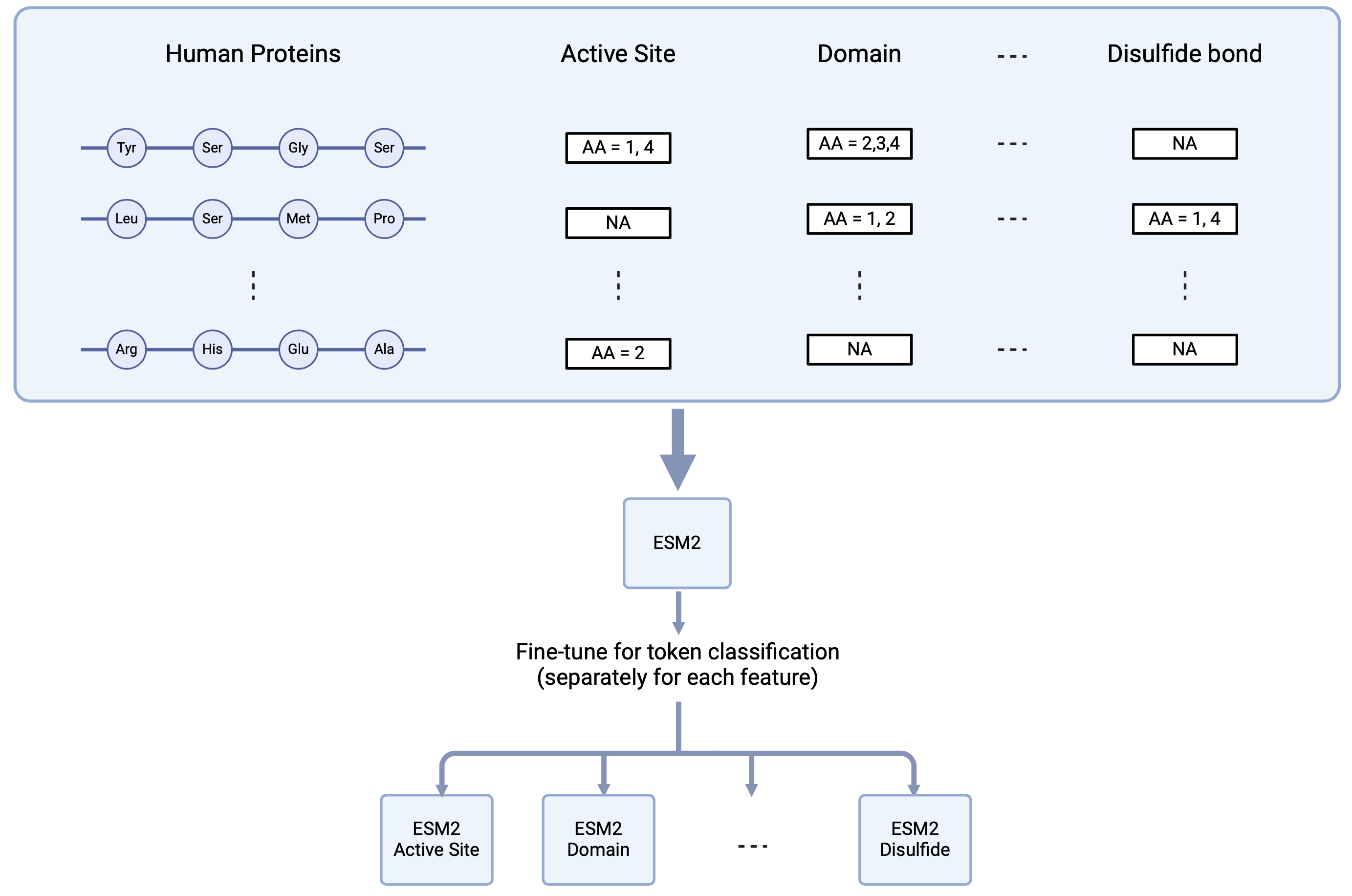} 
    \caption{\footnotesize Fine-tuning ESM2 for amino acid classification: 20,434 human protein sequences were downloaded from UniProtKB/SwissProt. Each sequence was annotated with 20 features at amino acid resolution. For each feature, ESM2 was fine-tuned to perform amino acid classification based on presence or absence of the feature. This resulted in 20 fine-tuned ESM2 models. AA, amino acid; NA, not available. Figure created with \url{BioRender.com}.}
    \label{fig_esm_finetune}
\end{figure}

\begin{figure}[h]
    \centering
    \includegraphics[width=0.8\textwidth]{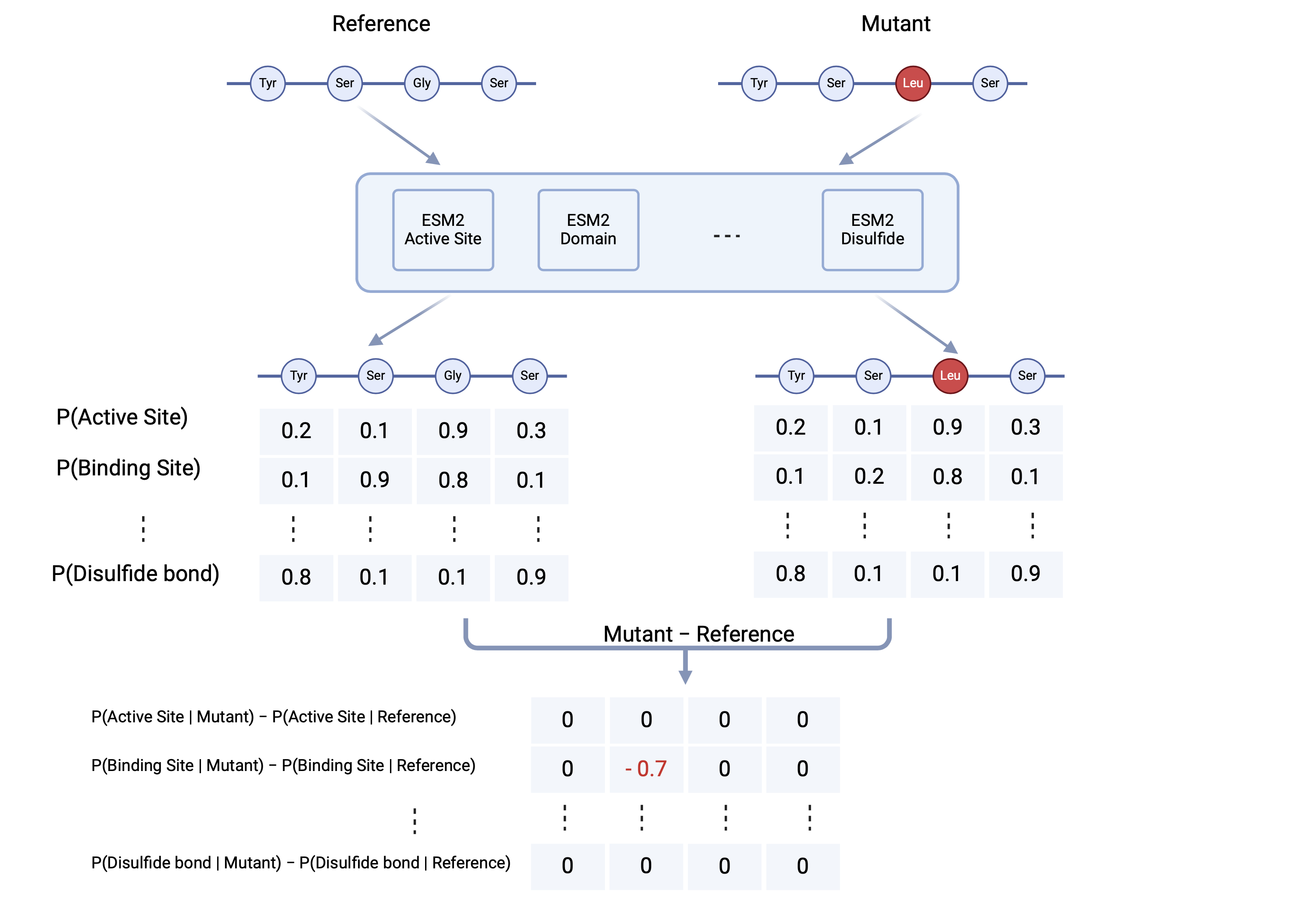} 
    \caption{\footnotesize Mechanistic variant interpretation using fine-tuned ESM2 models: to understand the impact of a missense variant, the reference and alternate protein sequences are analyzed by the ESM2 fine-tuned models. Then the difference between the prediction probabilities of alternate and reference features is calculated. Finally, a threshold is applied to detect gain or loss of features upon mutation. Figure created with \url{BioRender.com}.}
    \label{fig_diff_inference}
\end{figure}

To address these gaps, we harness the power of protein language models (PLMs), specifically ESM2 \cite{Lin2023}, for variant classification and interpretation. PLMs, such as ESM2, have demonstrated remarkable potential in capturing structural and functional properties of proteins through pretraining on large-scale protein sequence data \cite{Bepler2021}. Recent studies have shown the efficacy of fine-tuning ESM2 for various downstream tasks \cite{Schmirler2023,Schreiber2023}. In this study, we fine-tune the ESM2 models to assess various protein features at amino acid resolution (Figure \ref{fig_esm_finetune}), then utilize it to: 1) find features that are enriched in pathogenic vs. benign variants, which can be considered as critical functional regions, 2) compare the characteristics of proteins with reference or alternate alleles to understand how missense variants affect protein functionality (Figure \ref{fig_diff_inference}).

Furthermore, we demonstrate the practical application of our model by reclassifying VUS in the gnomAD database \cite{Chen2023}. By integrating our feature-based approach, we also provide protein- and feature-specific insights into how missense variants influence protein structure and function. This work not only contributes to improving variant interpretation but also offers a framework for leveraging PLMs to elucidate the functional impact of genetic variation.

\section{METHODS}
\label{methods}

\subsection{Data collection}

We selected 20,434 human proteins from UniProtKB/Swiss-Prot \cite{Boutet2007} and extracted their amino acid sequence as well as their protein family membership. We annotated the proteins with 20 features including:

\begin{itemize}
    \item Functional features: active site, binding site, and DNA binding site
    \item Sub-cellular location: topological domain and trans-membrane
    \item Post-transcriptional modification (PTM) and processing: disulfide bond, modified residue, propeptide, signal peptide, and transit peptide 
    \item Structure: $\beta$-strand, $\alpha$-helix, and turn
    \item family and domain: coiled coil, compositional bias, domain, motif, region, repeat, and zinc finger
\end{itemize}

\subsection{Fine-tuning and evaluation}

To create train, validation, and test splits, we clustered all protein sequences using MMseqs2 \cite{Steinegger2017} with thresholds of 20\% coverage and 20\% sequence identity. For each feature, the annotated proteins were divided into 70\% training, 15\% validation, and 15\% testing sets, ensuring minimal data leakage by performing the splits at the cluster level. Afterward, for each feature, we fine-tuned ESM2 with four model sizes—8 million, 35 million, 150 million, and 650 million parameters—to classify amino acids based on the presence or absence of the feature. We trained each model for 5 epochs, and kept the checkpoint with lowest validation loss. We used one Nvidia A100 GPU for training. We evaluated each model on their test set using $F_1$ scores, aggregated using macro averaging \cite{F1_macro}.

\subsection{Protein annotation inference}

For each feature, we extracted the amino acid sequences from all proteins that lacked information about that feature. We utilized the corresponding fine-tuned model to predict presence or absence of the feature at each amino acid. To check the quality of predictions, we compared the distribution of GERP conservation scores\cite{Cooper2005} and REVEL pathogenicity scores\cite{Ioannidis2016} between labeled and predicted amino acids.

\subsection{Applications}

\begin{itemize}

\item Variant reclassification: according to the ACMG/AMP guidelines \cite{Richards2015}, missense variants that are located in a mutational hot spot and/or critical functional domains are more likely to be pathogenic (moderate evidence of pathogenicity, PM1). To identify such regions, we obtained 46,504 missense pathogenic and 53,169 missense benign variants from ClinVar \cite{Landrum2013} (variants with conflicting classification were removed). We also extracted 18,991 non-redundant missense variants with minor-allele frequency $\geq$ 0.02 from gnomAD \cite{Chen2023}, and added them to the set of benign variants (they are considered benign due to high frequency in population). We performed two-sided Fisher's exact test to identify protein features that are significantly enriched in pathogenic or benign variants. After detecting regions with enrichment of pathogenic variants, we used them to reclassify variants of unkown significance (VUS) in gnomAD. To do so, we extracted all missense variants from gnomAD, and assigned a probability of pathogenicity (PoP) without using PM1. Then we focused on VUS and calculated a new PoP score by adding the PM1 evidence which is applied for missense variants located in regions with high enrichment of pathogenic variants. Finally, we calculated the fraction of VUS that were reclassified by adding PM1.

\item Variant interpretation: to understand the potential impact of a missense variant on the protein, we designed a workflow that can provide insight into the variant mechanism (Figure \ref{fig_diff_inference}). In summary, we pass the reference and mutant protein sequences into ESM2 fine-tuned models. Then we subtract the prediction probabilities of mutant features from reference features. To detect a gain or loss of a certain feature upon mutation, amino acid label must change and the absolute value of differential score should exceed a threshold (we chose 0.5 as the threshold but it is possible to have a lower threshold to increase sensitivity or a higher threshold to increase specificity). By doing so, we can predict the changes in protein features at amino acid resolution, which can be helpful for designing follow-up functional studies. To demonstrate this application, we downloaded 6,974 curated variants across 107 genes from ClinGen \cite{Rehm2015}, and kept genes with at least one pathogenic and one benign missense variants. Then we applied the variant interpretation workflow (Figure \ref{fig_diff_inference}) for all the selected variants.

\end{itemize}

\section{RESULTS}

\label{results}

\subsection{Fine-tuning and evaluation}

We retrieved amino acid sequences of 20,434 human proteins from UniProtKB/Swiss-Prot \cite{Boutet2007}, along with associated annotations. The number of annotated human proteins per feature in UniProtKB/Swiss-Prot is illustrated in Figure \ref{fig_s1}. To create train, validation, and test splits, we clustered all protein sequences using MMseqs2 \cite{Steinegger2017} with thresholds of 20\% coverage and 20\% sequence identity, resulting in 7,538 unique clusters. For each feature, the annotated proteins were divided into 70\% training, 15\% validation, and 15\% testing sets, ensuring minimal data leakage by performing the splits at the cluster level.

\begin{figure}[h]
    \centering
    \includegraphics[width=0.95\textwidth, height=0.4\textwidth]{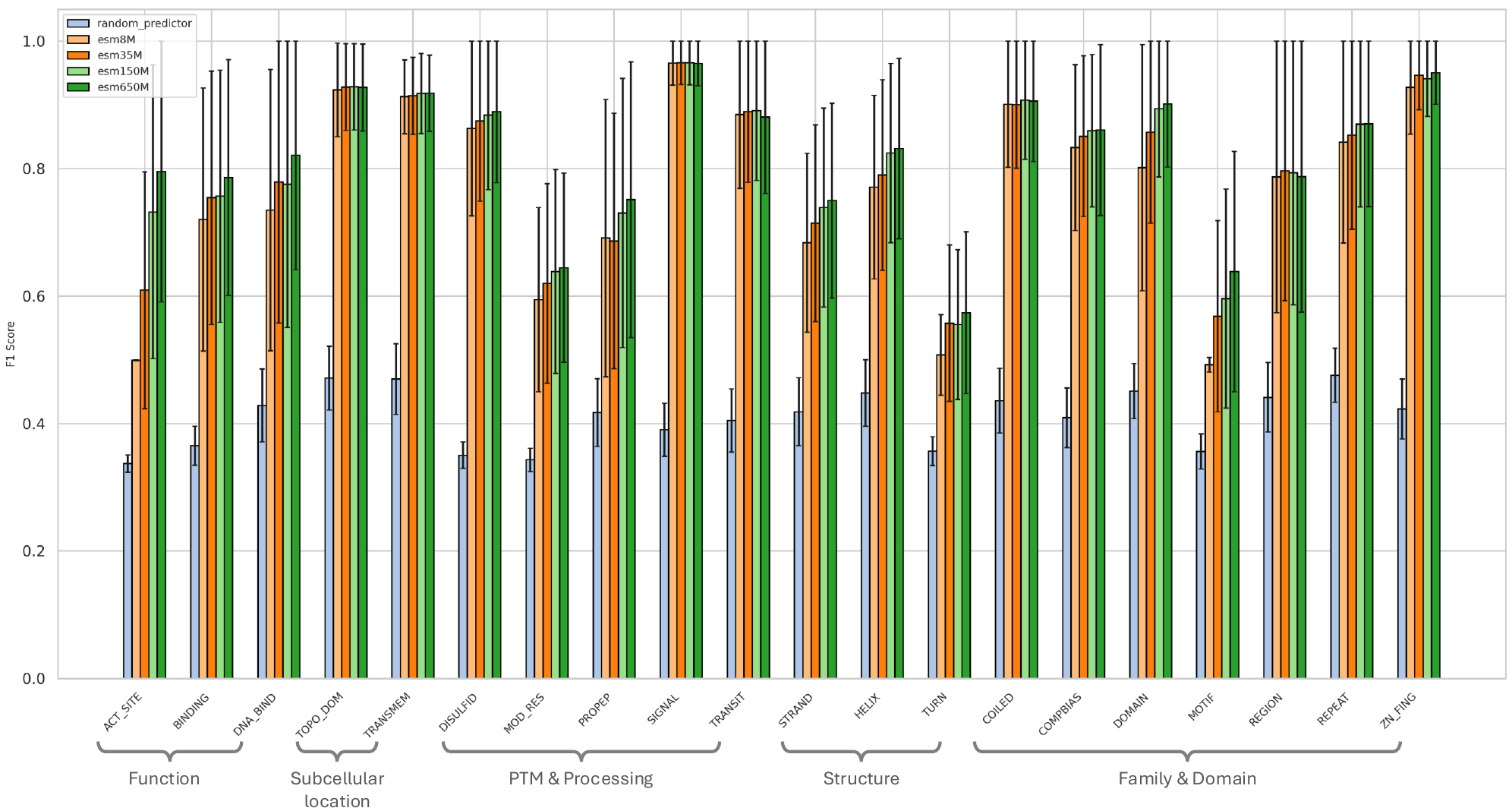} 
    \caption{\footnotesize Performance of fine-tuned models: Feature-specific test splits were used to evaluate the performance of the fine-tuned models. The vertical axis shows the $F_1$ scores, aggregated using macro averaging \cite{F1_macro}, and error bars indicate the standard deviation across all proteins in the test set.}
    \label{fig_metrics}
\end{figure}

We fine-tuned ESM2 with four model sizes: 8 million, 35 million, 150 million, and 650 million parameters. For each feature, the training split was used to fine-tune the ESM2 PLMs for amino acid classification, while the validation split was used to select the model checkpoint with the lowest validation loss. Model performance was evaluated on feature-specific test sets (Figure \ref{fig_metrics}), with a random predictor included as a baseline. All fine-tuned models outperformed the baseline, with performance generally improving as model size increased. Based on these results, we selected the 650-million-parameter model (esm650M) for subsequent analyses. Figure \ref{fig_s2} presents the precision, recall, and $F_1$ score of esm650M across all features.

\subsection{Protein annotation inference}

\begin{figure}[h]
    \centering
    \begin{minipage}{0.4\textwidth}
        \centering
        \includegraphics[width=\textwidth]{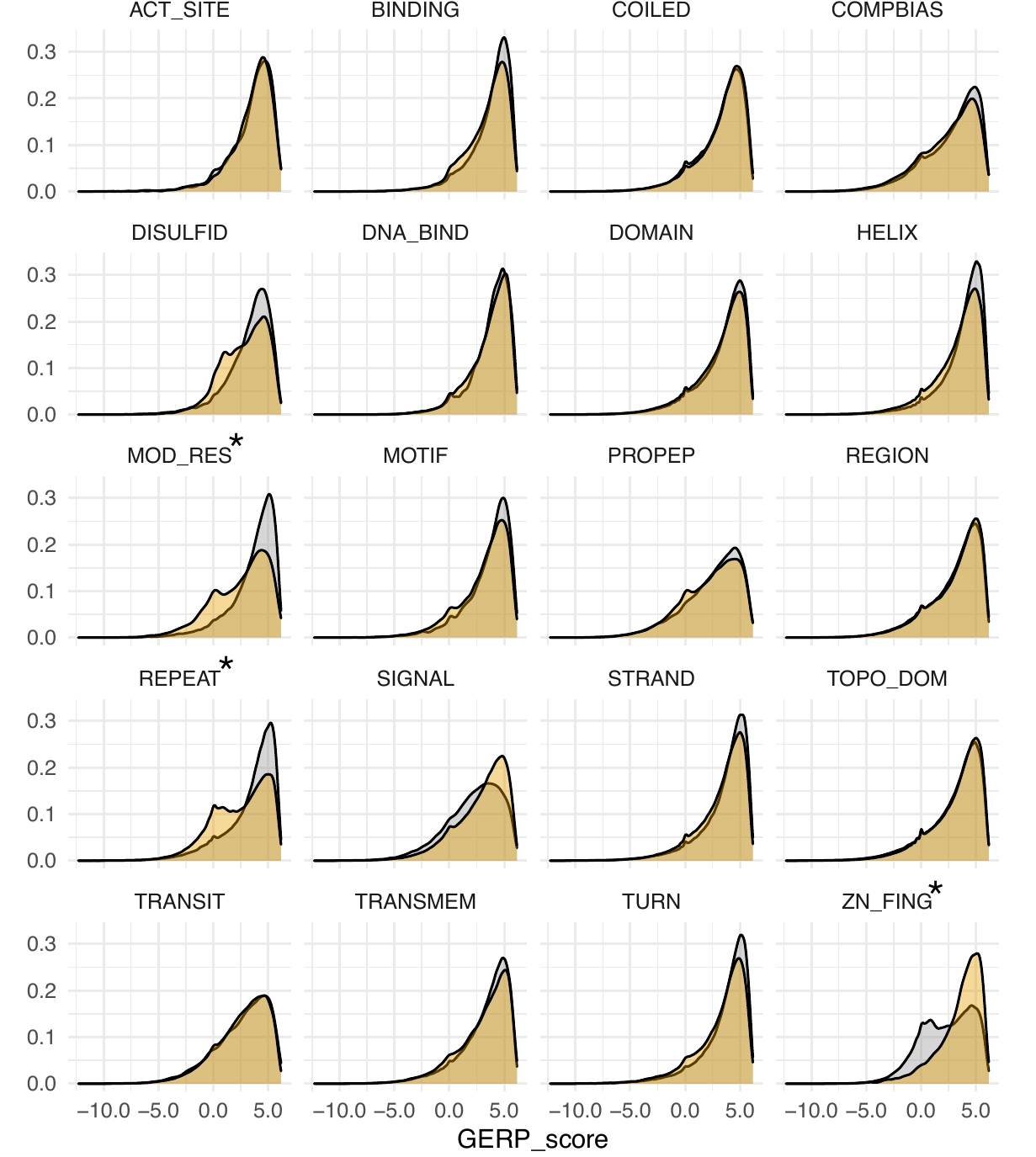}
    \end{minipage}
    \hspace{0.05\textwidth}
    \begin{minipage}{0.48\textwidth}
        \centering
        \includegraphics[width=\textwidth]{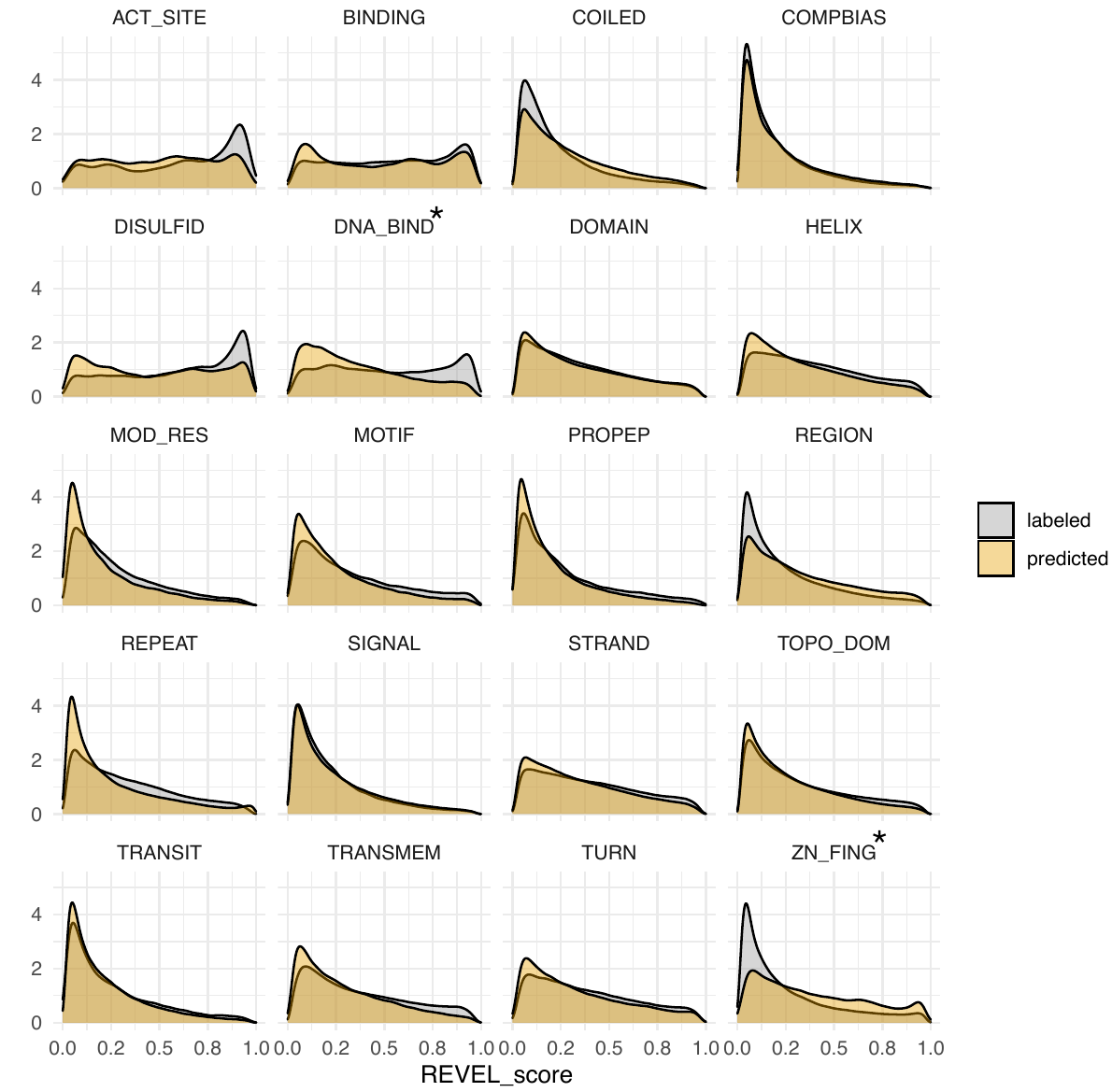}
    \end{minipage}
    \caption{\footnotesize Comparison of Characteristics Between Predicted and Labeled Amino Acids:
The left panel shows the distribution of conservation scores, while the right panel displays pathogenicity scores for labeled and predicted amino acids across all features. Features marked with an asterisk (*) indicate significant differences (p-value $\leq$ 0.05 and effect size $\geq$ 0.2) based on the Kolmogorov-Smirnov test.}
    \label{fig_gerp_revel}
\end{figure}

We utilized fine-tuned esm650M models to predict the presence or absence of features in proteins lacking annotations. The number of labeled and predicted proteins, as well as amino acids, is detailed in Figure \ref{fig_s3}. To evaluate prediction quality, we analyzed the distribution of conservation scores (GERP \cite{Cooper2005}) and variant pathogenicity scores (REVEL \cite{Ioannidis2016}) between labeled and predicted amino acids (Figure \ref{fig_gerp_revel}). Using the Kolmogorov-Smirnov test \cite{Karson1968-hq} to compare distributions per feature, we found that most features showed no significant differences. Notable exceptions included DNA binding sites and zinc fingers for REVEL scores, as well as modified residues, repeats, and zinc fingers for GERP scores.

\subsection{Applications}

\subsubsection*{1. Variant classification}

\begin{figure}[h]
    \centering
    \includegraphics[width=0.9\textwidth]{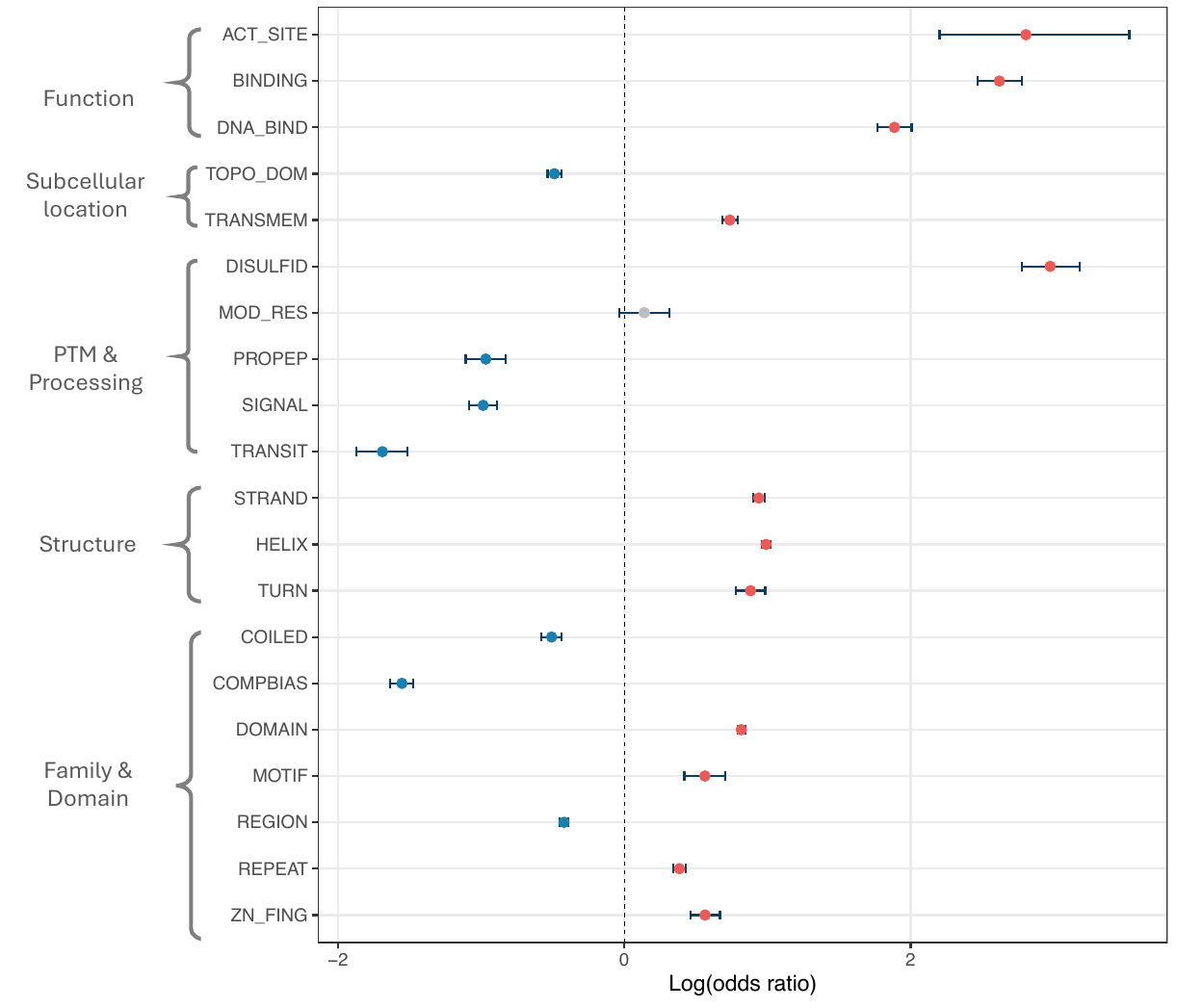} 
    \caption{\footnotesize Enrichment of features in pathogenic and benign variants: A two-sided Fisher's exact test was conducted for each feature using pathogenic variants from ClinVar and benign variants from ClinVar/gnomAD. The odds ratio (OR) was calculated as 
\ensuremath{OR = \frac{a \cdot d}{b \cdot c}}, 
where \ensuremath{a} and \ensuremath{b} are the number of pathogenic and benign variants, respectively, within a given feature, and \ensuremath{c} and \ensuremath{d} are the number of pathogenic and benign variants outside the feature. Twelve features were significantly enriched in pathogenic variants (red dots), while seven features were enriched in benign variants (blue dots). The modified residue feature showed no significant enrichment in either pathogenic or benign variants (grey dot).}

    \label{fig_OR}
\end{figure}

Using 46,504 pathogenic variants from ClinVar and 72,150 benign variants from ClinVar/gnomAD, we performed two-sided Fisher's exact tests to identify protein features significantly associated with pathogenic or benign variants. Figure \ref{fig_OR} highlights 12 features enriched in pathogenic variants, including active site, binding site, DNA binding site, transmembrane regions, disulfide bonds, $\beta$-strands, $\alpha$-helices, turns, domains, motifs, repeats, and zinc fingers. Leveraging these 12 features, we aimed to reclassify variants of uncertain significance (VUS) in gnomAD.

To detect VUS, we applied ACMG/AMP criteria (see Appendix \ref{appendix}) and calculated a probability of pathogenicity (PoP) for all missense variants in gnomAD. A total of 1,692,568 variants with $0.1 <$ PoP $< 0.9$ were classified as VUS. We then refined the PoP score by incorporating PM1 evidence, applied to missense variants located in the 12 protein features with significant enrichment of pathogenic variants. This refinement led to the reclassification of 110,304 (6.5\%) VUS as pathogenic.

\subsubsection*{2. Variant interpretation}

\begin{figure}[h]
    \centering
    \includegraphics[width=0.8\textwidth]{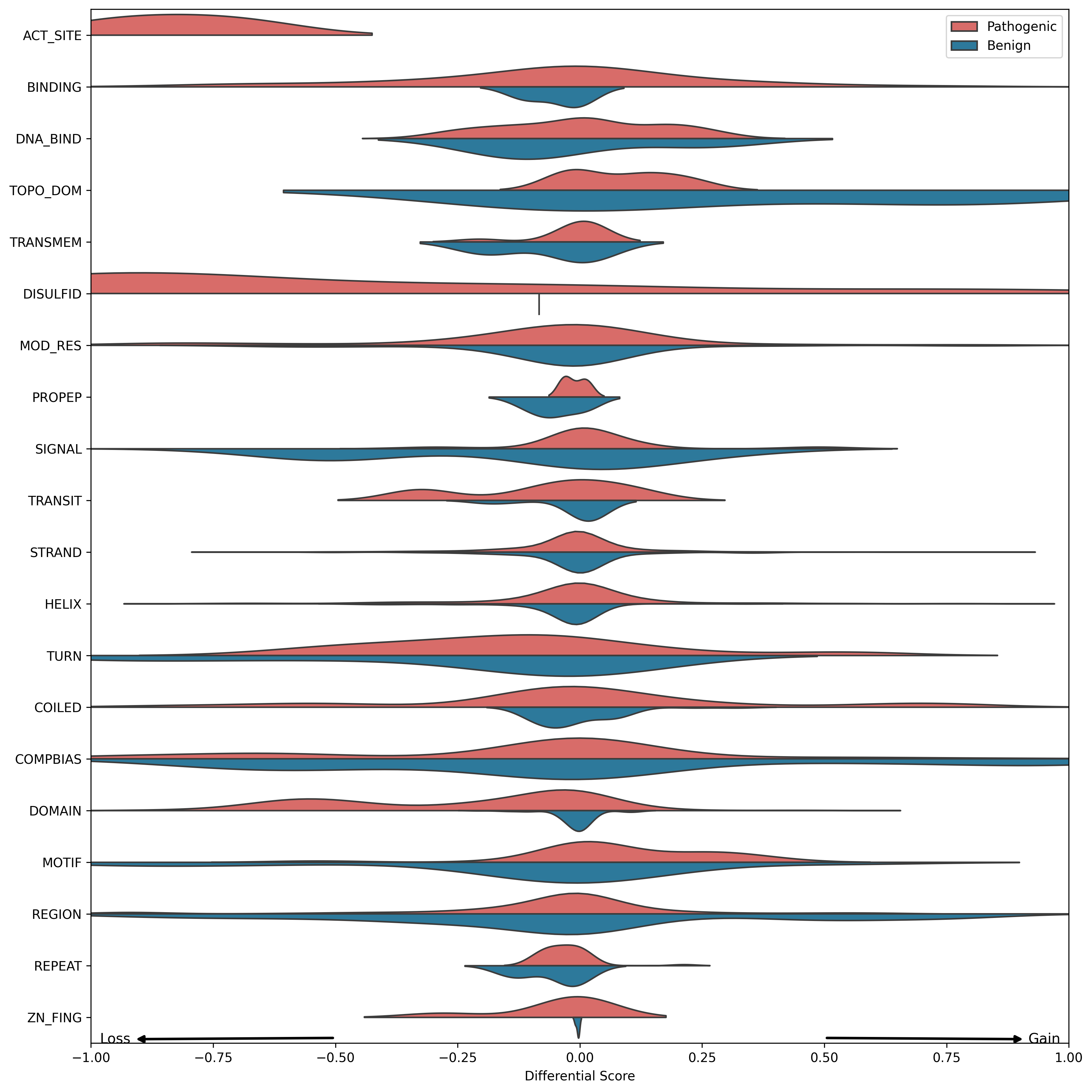} 
    \caption{\footnotesize Predicted impact of pathogenic and benign missense variants on protein features: A total of 771 curated variants across 54 genes, each with at least one pathogenic and one benign missense variant, were identified from ClinGen. The variant interpretation workflow (Figure \ref{fig_diff_inference}) was applied individually to each variant, and the differences in predicted probabilities between the alternate and wild-type proteins were recorded.}
    \label{fig_aachange_barplot}
\end{figure}

\begin{figure}[h]
    \centering
    \includegraphics[width=0.8\textwidth]{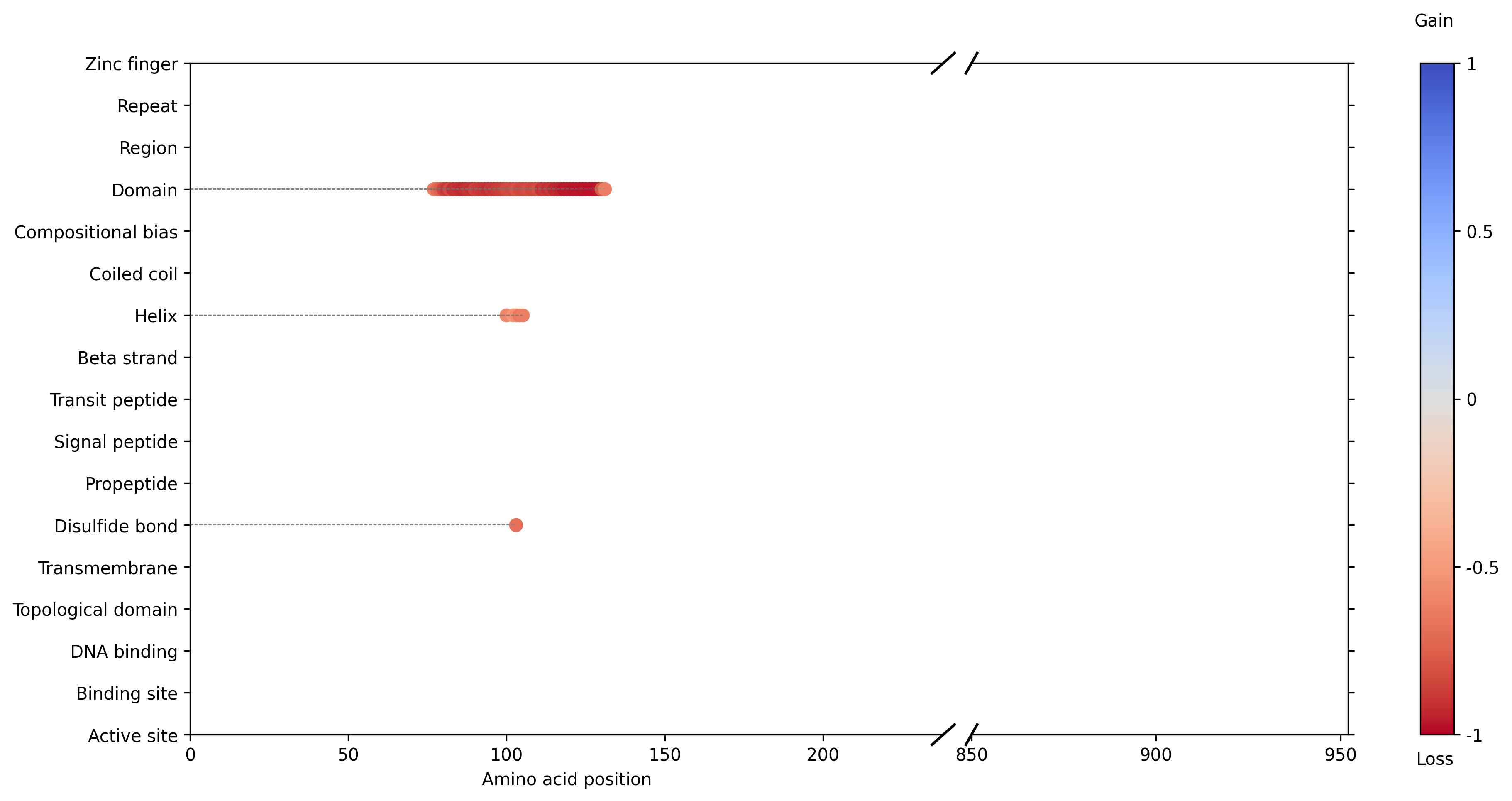} 
    \caption{\footnotesize Predicted impact of the p.Cys103Gly variant on \textit{GAA}: The variant interpretation workflow (Figure \ref{fig_diff_inference}) was used to predict changes in protein features caused by the \textit{GAA}:p.Cys103Gly variant. Each dot represents the predicted loss or gain of a feature at the corresponding position. The \textit{GAA}:p.Cys103Gly variant is predicted to result in the complete loss of the P-type domain in the mutant protein.}
    \label{fig_gaa}
\end{figure}

\begin{figure}[h!]
    \centering
    \includegraphics[width=0.8\textwidth]{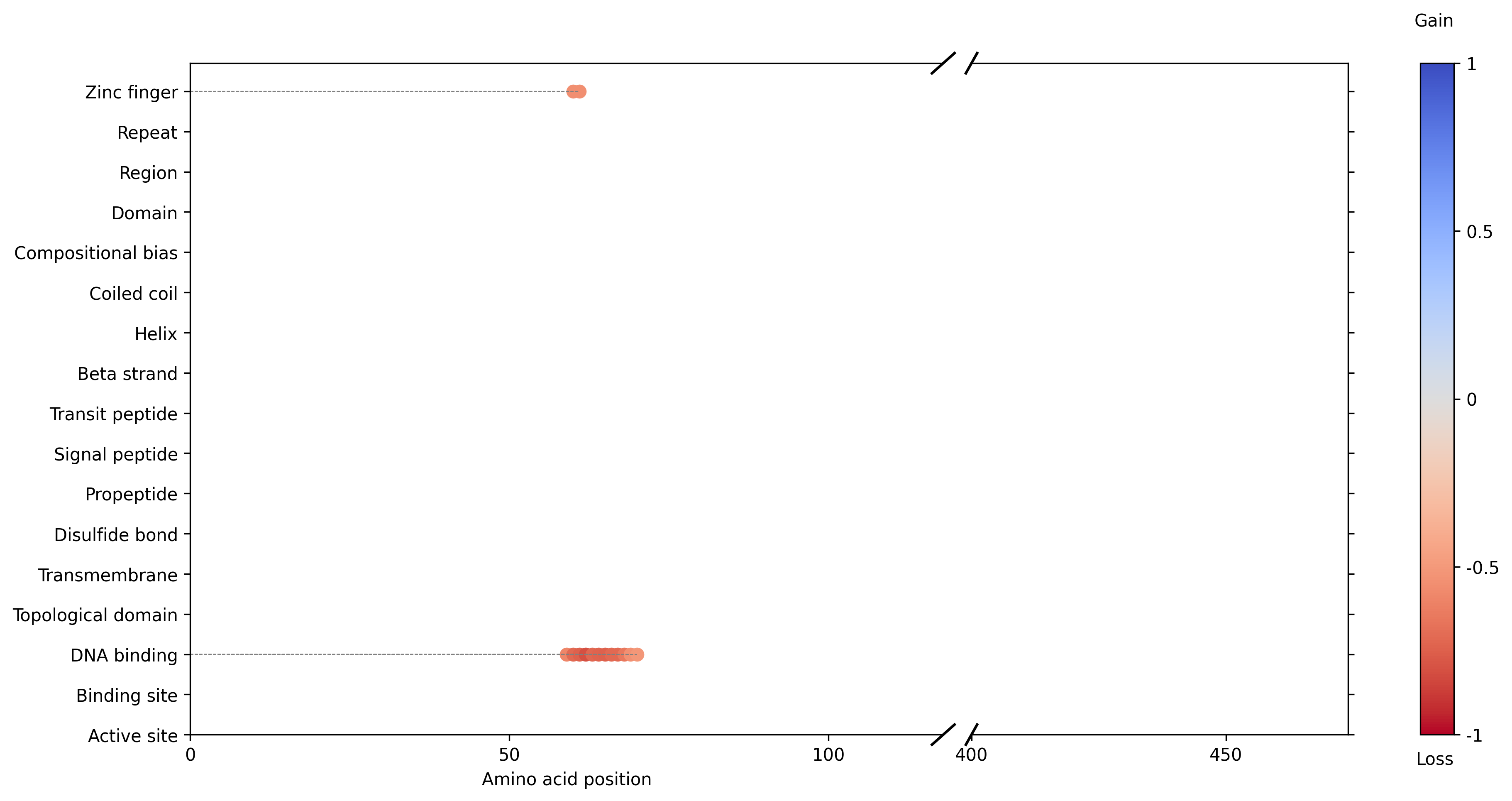} 
    \caption{\footnotesize Predicted impact of the p.Arg63Gln variant on \textit{HNF4A}: The variant interpretation workflow (Figure \ref{fig_diff_inference}) was used to predict changes in protein features caused by the \textit{HNF4A}:p.Arg63Gln variant. Each dot represents the predicted loss or gain of a feature at the corresponding position. The \textit{HNF4A}:p.Arg63Gln variant is predicted to cause a partial loss of the DNA-binding domain in the mutant protein.}
    \label{fig_hnf4a}
\end{figure}

We identified 771 curated variants in 54 genes that had at least one pathogenic and one benign missense variant from ClinGen \cite{Rehm2015}. To understand the potential impact of each missense variant, we utilized the workflow introduced in figure \ref{fig_diff_inference}. In summary, we passed the reference and alternate protein sequences through the fine-tuned esm650M, then calculated the difference between predicted probabilities of features at each amino acid. We observed that pathogenic missense variants impact certain protein features, especially loss of active sites, disulfid bonds, an functional domains (Figure \ref{fig_aachange_barplot}). Regarding benign variants, we observed that they have much milder effect, and that they can result in loss of compositional bias and signal (Figure \ref{fig_aachange_barplot}), meaning that proteins can tolerate mutations in these features, which is in agreement with our previous observation (Figure \ref{fig_OR}). To demonstrate the power of our method, we chose two pathogenic variants (\textit{GAA}:p.Cys103Gly and \textit{HNF4A}:p.Arg63Gln) with deleterious effect on protein domains and DNA binding site, respectively. We observed that \textit{GAA}:p.Cys103Gly results in the complete loss of P-type domain from the alternate protein, as well as loss of disulfide bonds and $\alpha$-helix (Figure \ref{fig_gaa}). Regarding \textit{HNF4A}:p.Arg63Gln, we could detect a partial loss of DNA binding domain as well as loss of the zinc fingers (Figure \ref{fig_hnf4a}).

\section{DISCUSSION}

This study introduces a novel application of the ESM2 protein language model to deepen our understanding of the functional consequences of missense variants. By fine-tuning ESM2 on specific protein features, we developed a robust and accessible toolset capable of classifying and interpreting missense variants with remarkable precision and detail.

Using these fine-tuned models, we quantified the enrichment of pathogenic variants across protein features and successfully reclassified 6.5\% of variants of uncertain significance (VUS) in gnomAD as pathogenic. Furthermore, we demonstrated the models’ ability to predict structural and functional changes caused by missense variants. For instance, our analysis identified the loss of the P-type domain in \textit{GAA} and the partial loss of the DNA-binding domain in \textit{HNF4A}, directly linking these alterations to disease mechanisms.

While our models are effective and accessible, there remains a critical need for greater interpretability in machine learning tools \cite{Chen2024}. Currently, many advanced models operate as black boxes, providing accurate predictions without offering insights into how those predictions are made. This lack of transparency can hinder trust and adoption among biologists and clinicians, who require clear explanations of model outputs to validate findings and design follow-up experiments. Developing more interpretable models will enable users to better understand the underlying features driving variant classification, ultimately increasing the utility of these tools in research and clinical settings \cite{Jnes2024,Saadat_2024_DNALM,saadat2024_MOI}.

The accessibility of our current models ensures that biologists and clinicians, even those without expertise in machine learning, can readily use them to explore the molecular basis of missense variants. However, further advancements in interpretability will make these tools even more actionable, facilitating their integration into workflows for functional studies and patient care.

Despite these advancements, certain limitations persist. Some protein features were excluded due to their suboptimal contribution to model performance, underscoring the inherent complexity and variability of these features. Additionally, while our models perform well in annotated protein regions, their predictions for less-characterized regions require further experimental validation to ensure reliability. In terms of variant datasets, ClinVar remains one of the highest-quality resources available for our analysis, despite its bias toward clinically relevant genes. Conversely, gnomAD offers broader coverage with less bias overall, but it still exhibits an enrichment for European populations.

Overall, this study confirms the transformative potential of machine learning in genomics which can be used together with other tools to improve our understanding of genetic diseases \cite{saadat2025NMD,saadat2025zygosity,ruiz2025benchmarking}. By improving variant interpretation and addressing challenges in model interpretability, our approach could significantly contribute to personalized medicine, enabling advancements in patient diagnosis, risk assessment, and treatment planning.


\section*{RESOURCE AVAILABILITY}


\subsection*{Lead contact}


Requests for further information and resources should be directed to and will be fulfilled by the lead contact, Jacques Fellay (jacques.fellay@epfl.ch).

\subsection*{Data and code availability}


The code for this study is available \href{https://github.com/AliSaadatV/ESM2-Missense-Impact-Analysis}{here}. All the fine-tuned models can be accessed \href{https://huggingface.co/collections/AliSaadatV/esm2-finetuned-models-669129d121b55573425a869b}{here}. 

\section*{ACKNOWLEDGMENTS}


This work was funded by the Swiss National Science Foundation via grant \#197721 and by the Swiss State Secretariat for Education, Research and Innovation via contribution to project "UNDINE", SBFI No. 23.00322.

\section*{AUTHOR CONTRIBUTIONS}


Conceptualization, A.S.; methodology, A.S..; investigation, A.S., and J.F.; writing-–original draft, A.S.; funding acquisition, J.F.; supervision, J.F.

\section*{DECLARATION OF INTERESTS}


The authors declare no competing interests.

\section*{DECLARATION OF GENERATIVE AI AND AI-ASSISTED TECHNOLOGIES}


During the preparation of this work, the authors used GPT-4 in order to improve writing and readability. 

\newpage


\bibliography{references}

\bigskip


\newpage

\section{Supplementary figures}

\newcounter{suppfigure}
\renewcommand{\thefigure}{S\arabic{suppfigure}} 
\setcounter{suppfigure}{1}
\renewcommand{\figurename}{Supplementary Figure}

\begin{figure}[h]
    \centering
    \includegraphics[width=0.9\textwidth]{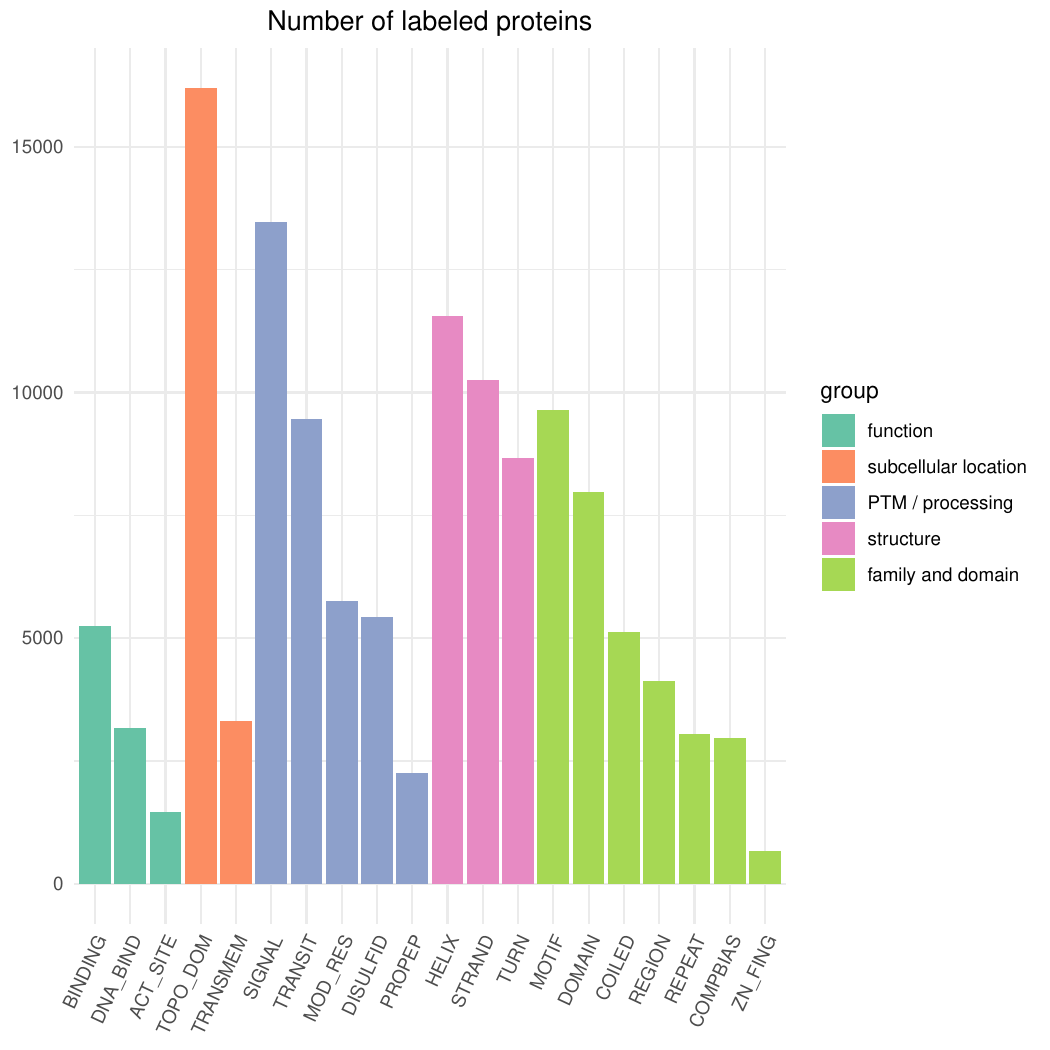} 
    \caption{\footnotesize Number of annotated protein per feature in UniProtKB/Swiss-Pro}
    \label{fig_s1}
    \refstepcounter{suppfigure}
\end{figure}

\newpage

\begin{figure}[h]
    \centering
    \includegraphics[width=0.9\textwidth]{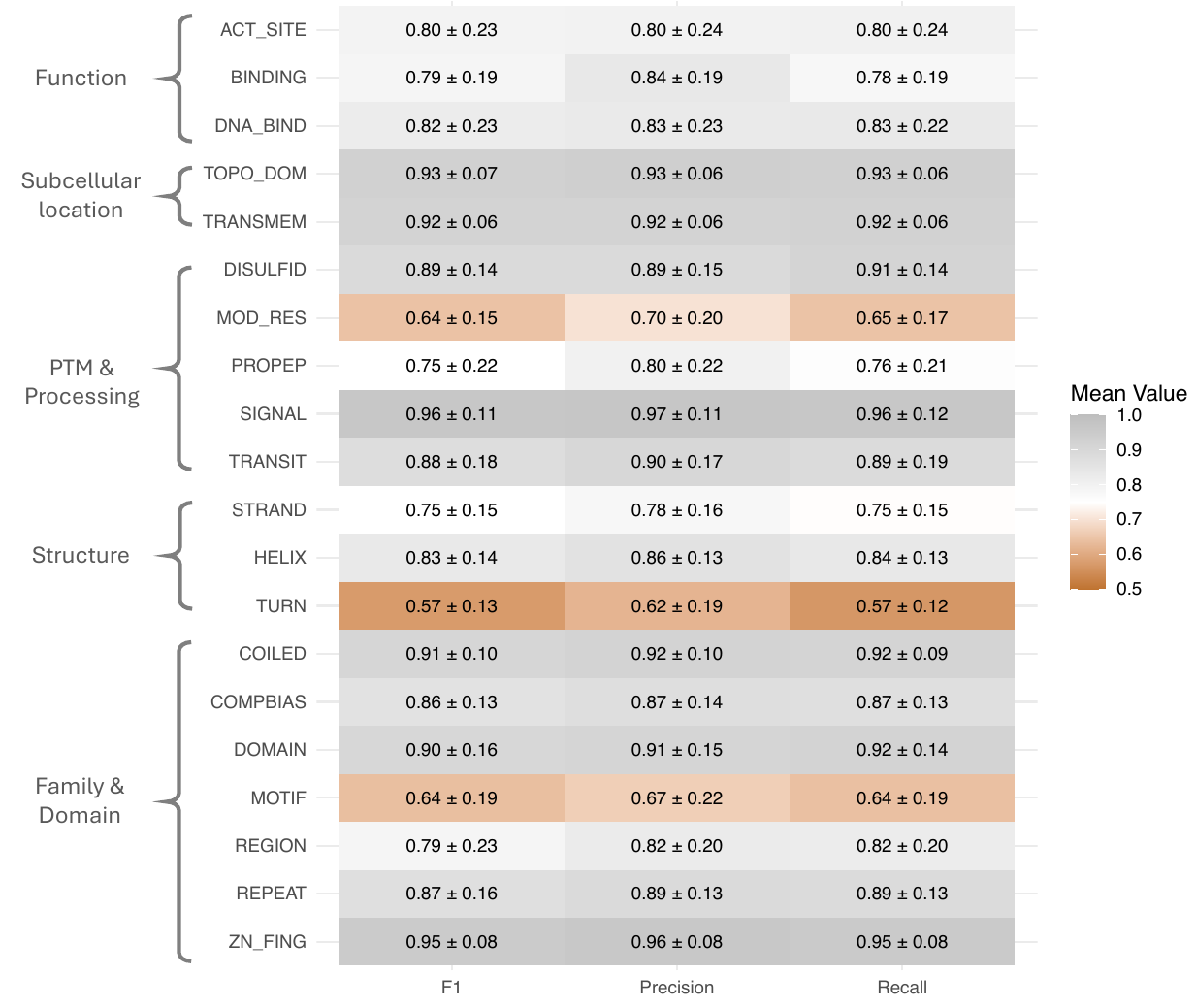} 
    \caption{\footnotesize Precision, recall, and $F_1$ of the fine-tuned esm650M model }
    \label{fig_s2}
    \refstepcounter{suppfigure}
\end{figure}

\newpage

\begin{figure}[h]
    \centering
    \includegraphics[width=0.9\textwidth]{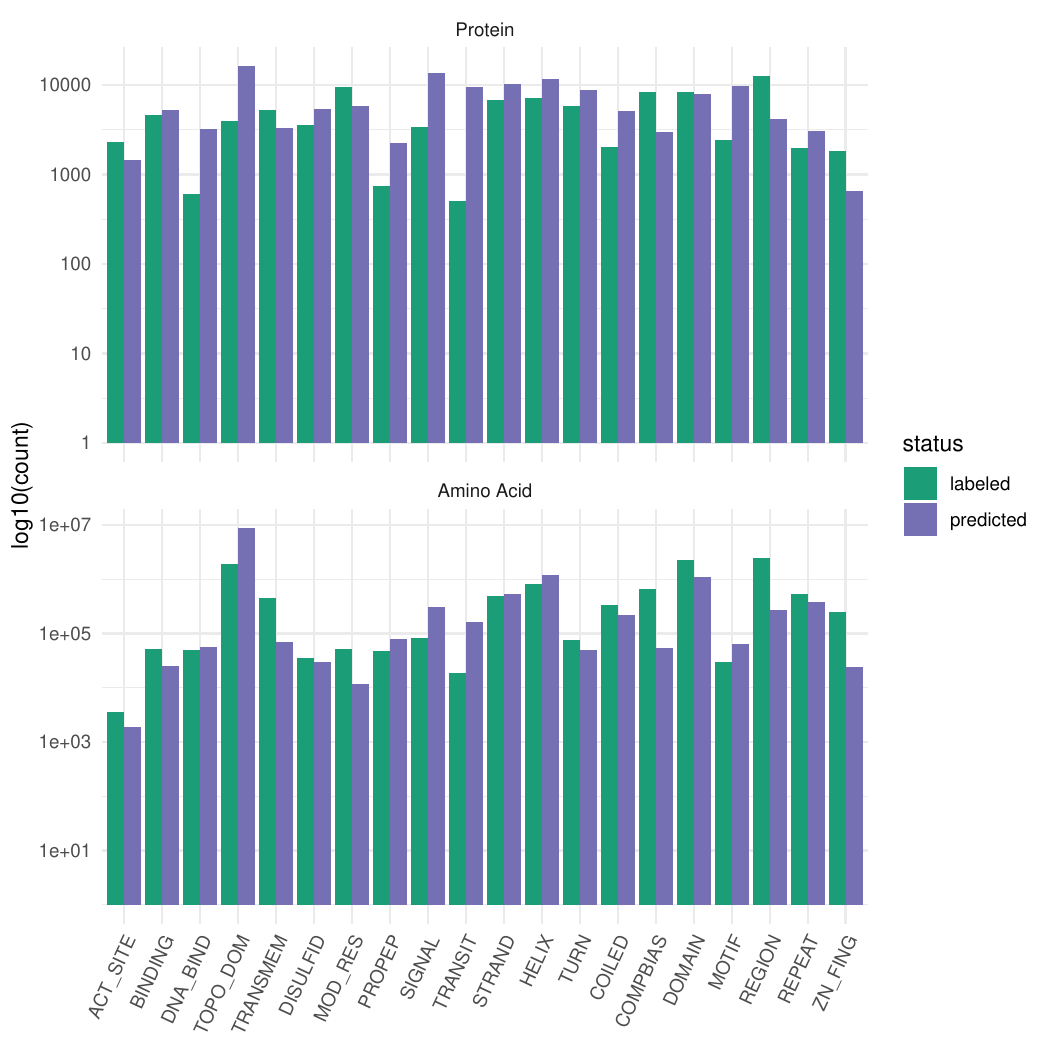} 
    \caption{\footnotesize Number of predicted and labeled proteins and amino acids}
    \label{fig_s3}
    \refstepcounter{suppfigure}
\end{figure}

\newpage

\appendix
\section{Appendix}
\label{appendix}

\subsection*{Probability of pathogenicity (PoP) calculation}

\begin{figure}[h]
    \centering
    \includegraphics[width=0.9\textwidth]{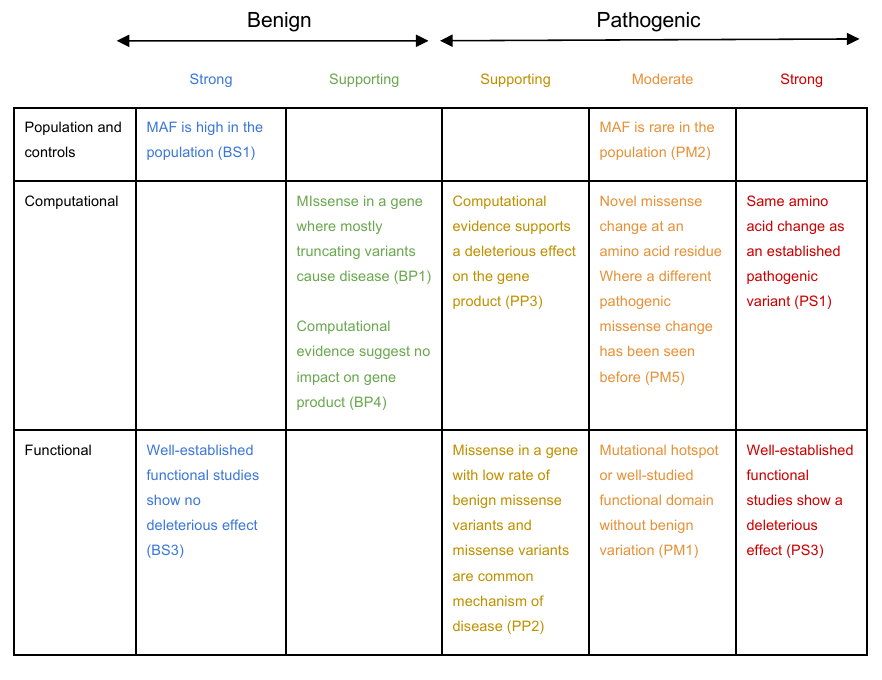} 
    \caption{\footnotesize the summary of ACMG/AMP criteria used for variant classification. MAF: minor allele frequency}
    \label{tab:tab1}
\end{figure}

We used ACMG/AMP guidelines \citep{Richards2015} to classify the variants into putative pathogenicity groups, as described in our previous works \citep{Saadat2023,saadat-fellay-2024-dna}. In summary, we gather all the available evidences for a variant. Figure \ref{tab:tab1} summarizes all the ACMG/AMP criteria that we used.

To calculate the probability of pathogenicity (PoP), we use the Bayesian framework developed by \citet{Tavtigian2018}. For a given variant, the PoP is calculated as follow:

\begin{center}
$P_x = \text{number of pathogenic criteria applied at the level of } x$ \\
$x \in \{\text{Strong}, \text{Moderate}, \text{Supporting}\}$ \\
[1em]

$B_y = \text{number of benign criteria applied at the level of } y$ \\
$y \in \{\text{Strong}, \text{Supporting}\}$ \\
[1em]

$\text{odds of pathogenicity (OP)} = 350^{(\frac{P_{\text{Strong}}}{2} + \frac{P_{\text{Moderate}}}{4} + \frac{P_{\text{Supporting}}}{8} - \frac{B_{\text{Strong}}}{2} - \frac{B_{\text{Supporting}}}{8})}$ \\
[1em]

$\text{probability of pathogenicity (PoP)} = \frac{OP \times 0.1}{((OP - 1) \times 0.1 + 1)}$
\end{center}

\end{document}